\documentclass[jgr,draft]{agujournal2019}
\usepackage{url} 
\usepackage[inline]{trackchanges} 
\usepackage{soul}
\draftfalse

\journalname{JGR: Space Physics}

\begin{document}

%
%


\title{Multi-scale observation of magnetotail reconnection onset: 2. microscopic dynamics}

%
%




\authors{	Kevin J. Genestreti\affil{1},
		Charles J. Farrugia\affil{2},
		San Lu\affil{3}\thanks{1},
		Sarah K. Vines\affil{4},
		Patricia H. Reiff\affil{5},
	      	Tai Phan\affil{6},
		Daniel N. Baker\affil{7},
		Trevor W. Leonard\affil{7}\thanks{4},
		James L. Burch\affil{8},
		Samuel T. Bingham\affil{4}\thanks{2},
		Ian J. Cohen\affil{4},
		Jason R. Shuster\affil{9}\thanks{3},
		Daniel J. Gershman\affil{9},
		Christopher G. Mouikis\affil{2},
		Anthony J. Rogers\affil{2},
		Roy B. Torbert\affil{1,2},
		Karlheinz J. Trattner\affil{7},
		James M. Webster\affil{5,8},
		Li-Jen Chen\affil{9},
		Barbara L. Giles\affil{9},
		Narges Ahmadi\affil{6},
		Robert E. Ergun\affil{6},
		Christopher T. Russell\affil{3},
		Robert J. Strangeway\affil{3},
		Rumi Nakamura\affil{10},
		Drew L. Turner\affil{4}}


\affiliation{1}{Earth Oceans and Space, Southwest Research Institute, Durham, New Hampshire, USA}
\affiliation{2}{Earth Oceans and Space, University of New Hampshire, Durham, New Hampshire, USA}
\affiliation{3}{Institute for Geophysics and Planetary Physics, University of California Los Angeles, Los Angeles, California, USA}
\affiliation{4}{Applied Physics Laboratory, Johns Hopkins University, Laurel, Maryland, USA}
\affiliation{5}{Rice Space Institute, Rice University, Houston, Texas, USA}
\affiliation{6}{Space Science Laboratory, University of California Berkeley, Berkeley, California, USA}
\affiliation{7}{Laboratory for Atmospheric and Space Physics, University of Colorado Boulder, Boulder, Colorado, USA}
\affiliation{8}{Space Science and Engineering Division, Southwest Research Institute, San Antonio, Texas, USA}
\affiliation{9}{Goddard Space Flight Center, National Aeronautics and Space Administration, Greenbelt, Maryland, USA}
\affiliation{10}{Space Research Institute, Austrian Academy of Sciences, Graz, Austria}
\thanks{1}{Now at School of Earth and Space Sciences, University of Science and Technology of China, Hefei, Anhui 230026, China}
\thanks{2}{Deceased}
\thanks{3}{Now at Earth Oceans and Space, University of New Hampshire, Durham, New Hampshire, USA}
\thanks{4}{Now at Cooperative Institute for Research in Environmental Sciences, CU, Boulder, and National Oceanic and Atmospheric Administration National Centers for Environmental Information}





\correspondingauthor{Kevin J. Genestreti}{kevin.genestreti@swri.org}




\begin{keypoints}
\item Magnetotail reconnection onset was triggered by electron tearing during a solar wind pressure pulse
\item Onset was characterized by the rapid collapse of the current sheet thickness and kinetic-scale flux rope formation
\item A primary X-line was established within minutes of the onset
\end{keypoints}

%
%

%
%


\begin{abstract}
We analyze the local dynamics of magnetotail reconnection onset using Magnetospheric Multiscale (MMS) data. In conjunction with MMS, the macroscopic dynamics of this event were captured by a number of other ground and space-based observatories, as is reported in a companion paper. We find that the local dynamics of the onset were characterized by the rapid thinning of the cross-tail current sheet below the ion inertial scale, accompanied by the growth of flapping waves and the subsequent onset of electron tearing. Multiple kinetic-scale magnetic islands were detected coincident with the growth of an initially sub-Alfv\'enic, demagnetized tailward ion exhaust. The onset and rapid enhancement of parallel electron inflow at the exhaust boundary was a remote signature of the intensification of reconnection Earthward of the spacecraft. Two secondary reconnection sites are found embedded within the exhaust from a primary X-line. The primary X-line was designated as such on the basis that (1) while multiple jet reversals were observed in the current sheet, only one reversal of the electron inflow was observed at the high-latitude exhaust boundary, (2) the reconnection electric field was roughly 5 times larger at the primary X-line than the secondary X-lines, and (3) energetic electron fluxes increased and transitioned from anti-field-aligned to isotropic during the primary X-line crossing, indicating a change in magnetic topology. The results are consistent with the idea that a primary X-line mediates the reconnection of lobe magnetic field lines and accelerates electrons more efficiently than its secondary X-line counterparts.
\end{abstract}

%
%

%


%
%
%
%

\section{Introduction}

Magnetic reconnection is a universal process in plasmas that dissipates magnetic energy in thin current sheets. Reconnection is among the most critical process for eruptive space weather events; it triggers solar flares and coronal mass ejections and enables the entry of solar wind mass and energy into Earth's magnetosphere \cite{Biskamp.2000}. In the magnetotail, reconnection powers geomagnetic storms and substorms \cite{Baker.1996,Angelopoulos.2008b}. In-situ observations of ongoing magnetotail reconnection have revealed a wealth of information about how reconnection operates. However, the onset mechanism is still under debate \cite{Sitnov.2019a}. The persistence of the onset mystery is simple to understand; reconnection onset is catalyzed by processes occurring at and between global and kinetic scales, which are difficult to resolve simultaneously, both observationally and in simulations. Understanding the processes that trigger reconnection in thin current sheets is one of the primary objectives of the Magnetospheric Multiscale (MMS) mission \cite{Burch.2016a}. 

It is widely acknowledged that current sheets must be thinned to the plasma-kinetic scale before reconnection occurs; however, not all thin current sheets reconnect. Non-zero magnetic fields at the current sheet center likely stabilize thin current sheets against reconnection \cite{BuchnerandZelenyi.1987,Pellat.1991}. The equatorial near-Earth magnetotail typically has $B_Z>0$. Magnetic and/or plasma thermal pressure gradients may play a critical role in stabilizing extended thin current sheets. Likewise, anisotropic ion and/or electron pressures may stabilize extended thin current sheets. Reconnection onset may be induced internally and/or externally. In Earth's magnetotail, external forcing by the solar wind may thin the cross-tail current layer down to the electron-kinetic scale, at which point the electron tearing instability and reconnection are initiated by electron non-gyrotropic pressure forces \cite{Liu.2014,Lu.2020,Lu.2022}. Time-dependent solar wind forcing may also be critical for precipitating the loss of tail current sheet equilibria, acting as an external trigger of reconnection \cite{BirnandSchindler.2002}. Internal trigger mechanisms include the spontaneous growth of thin current sheet instabilities, e.g., (1) the spontaneous magnetic flux release resulting from tailward gradients in the normal magnetic field ($\partial B_Z/\partial X<0$) \cite{MerkinandSitnov.2016} and (2) the ballooning interchange instability resulting from buoyant motion of magnetic flux tubes. These instabilities are purportedly responsible for thinning the current sheet down to the kinetic scale, leading to the growth of electron tearing and reconnection onset. The reader is directed to the recent reviews of \citeA{Sitnov.2019a} and \citeA{Pucci.2020} for further information.

Reconnection onset is characterized by the formation of one or more X-lines in the cross-tail current sheet. Reconnection may develop without the full and immediate participation of ions \cite{Lu.2020,Lu.2022}, in an initial electron-only mode. Reconnection is initially thought to precede slowly as low-Alfv\'en-speed flux tubes in the central plasma sheet are consumed and ejected, while the rate increases substantially after reconnection gains access to high-Alfv\'en-speed flux tubes in the lobes \cite{Hones.1977}. In the multiple X-line picture, not all X-lines need necessarily have identical reconnection rates \cite{Schindler.1974}; one primary X-line may gain access to higher latitude, higher Alfv\'en speed flux tubes than its neighboring X-lines. In this scenario, more slowly reconnecting secondary X-lines become entrained within the exhaust of their dominant counterpart and are expelled Earthward or tailward \cite{Eastwood.2005}. The delineation between (faster) primary and (slower) secondary X-lines also likely has implications for electron heating and the acceleration of suprathermals, given that the rates depend on $V_{Ai}^2$ \cite{Phan.2013a} and $|\vec{E}|$ \cite{Oka.2022}, respectively, both of which are larger for lobe reconnection (here, $V_{Ai}$ and $|\vec{E}|$ are the ion Alfv\'en speed and electric field magnitude, respectively).

This study investigates the onset of a magnetotail reconnection event in two papers; the first paper (hereafter Paper 1) uses a large amount of data from space and ground-based observatories to investigate the macroscopic physics of current sheet thinning. This paper uses multi-point data from MMS to investigate the kinetic physics of reconnection onset in the thinned and stretched current sheet. 

\section{Data and analysis methods}

MMS is a constellation of four spacecraft that fly in a tight, electron-scale tetrahedron. MMS is uniquely equipped with plasma and field instruments that make very high resolution measurements required to investigate electron-scale physics \cite{Burch.2016a}. During this study's event, which was on 2017 July 3, data from only three of the spacecraft are available. This study uses ion and electron measurements from the fast plasma investigation (FPI) \cite{Pollock.2016}. High-resolution burst-mode data from FPI are available at 30 and 150 milliseconds for electrons and ions (respectively). In fast survey mode, electron and ion data are both available at 4.5 seconds. Magnetic field data from the fluxgate magnetometers \cite{Torbert.2016a} are available at 128 Hz in burst mode and 8 Hz in survey mode; magnetic fields are calibrated to within $\leq0.1$ nT. Energetic electron data are obtained by the Fly's Eye Energetic Particle Sensor (FEEPS) \cite{Mauk.2016}. Electric field data from the electric field double probes instruments \cite{Ergun.2016a,Lindqvist.2016} are available at 8192 Hz in burst mode and 32 Hz in survey mode; electric fields are calibrated to within $\leq0.5$ mV/m. Mass-per-charge-resolved ion fluxes from the hot plasma composition analyzer \cite{Young.2016} were used to determine the ion composition of the plasma sheet; in this event, protons dominated, with O+ representing $\leq4\%$ of the total mass density. O+ is therefore neglected, and the plasma is assumed to be protons and electrons only throughout the paper. High-resolution burst-mode data are used whenever available, however they are not available during the entire period considered here. Burst and survey mode data are spliced together; we will point out which mode is being used whenever it is relevant.

Timing analysis is used to determine the speed and size of flux ropes. With 3-point data, full 3-d flux rope velocities cannot be calculated from timing analysis alone; however, if it is assumed that flux ropes propagate in the direction of the exhaust in which they are embedded, then only 2 spacecraft are needed to calculate the speed. Here, we use the third spacecraft for a consistency check. As in paper 1, wherein more details can be found, we approximate the time-dependent current sheet thickness using a 1-d Harris model. Briefly, the thickness of a Harris current sheet is determined by an asymptotic magnetic field strength and a maximum current density, which can be both determined remotely using measurements of the thermal pressures, magnetic field, and current density at the MMS location and assuming vertical pressure balance. Particle moments are used to calculate the current density. Plasma ion and electron pressure tensors are used to determine the species-dependent non-gyrotropy, which is quantified using the $\sqrt{Q}$ parameter \cite{Swisdak.2016}.

\section{Results}

\subsection{Event overview}

\begin{figure}
\noindent\includegraphics[width=39pc]{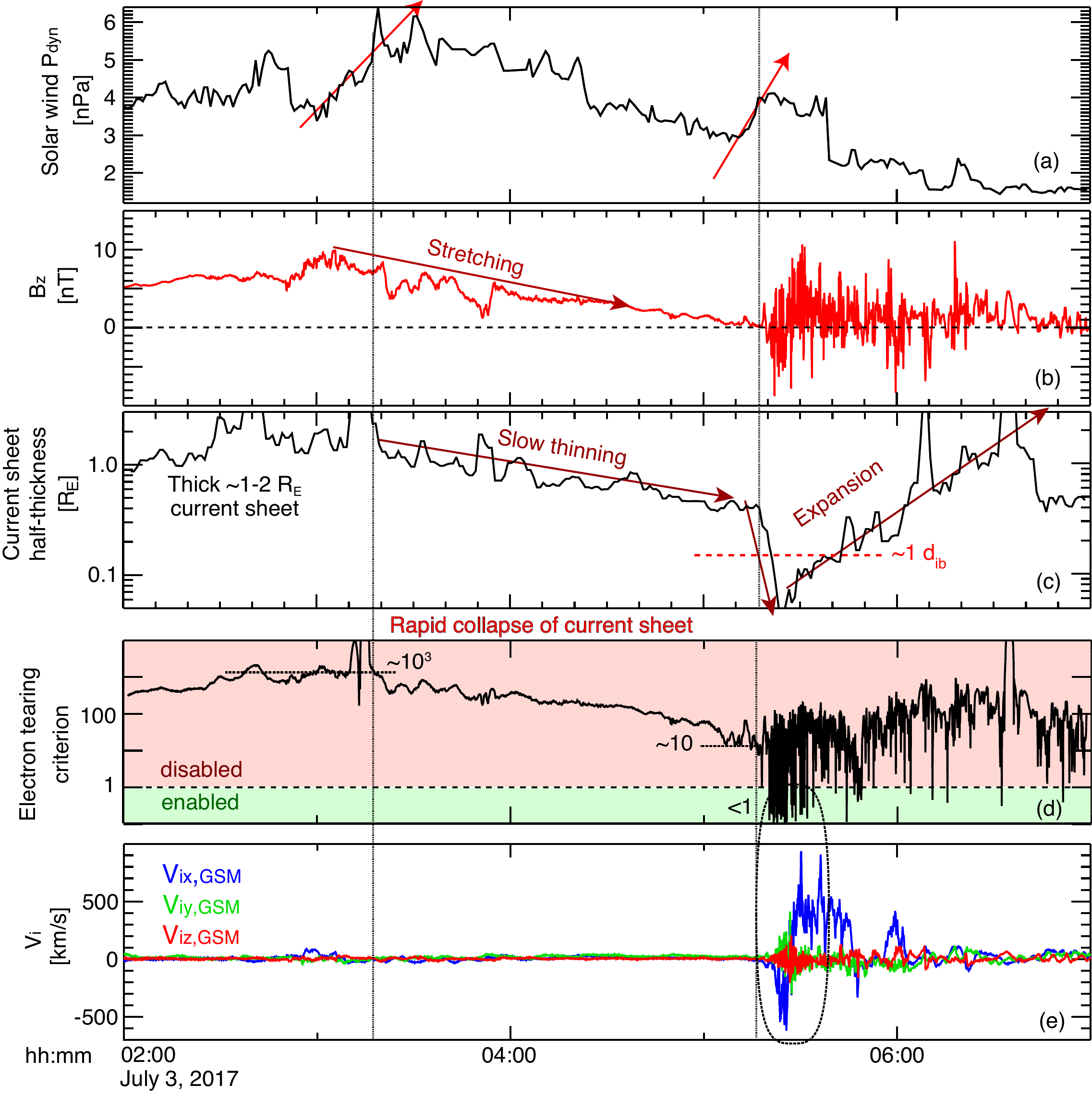}
\caption{(a) The solar wind dynamic pressure, measured by ACE at the Earth-sun L1 point and propagated to Earth's bow shock, (b) the north-south component of the magnetic field measured by MMS, (c) the current sheet thickness determined from MMS using the Harris equilibrium equation (see Paper 1), (d) the electron tearing criterion of \cite{Liu.2014}, determined from MMS (see text and equation 1), and (e) the ion bulk velocity measured by MMS. Vertical dashed lines mark the times of two transient enhancements in the solar wind dynamic pressure.}
\label{tearing}
\end{figure}

Figure \ref{tearing}a-c summarizes the evolution of the magnetotail, as observed by MMS, during the interval investigated by Paper 1. A first transient, 30-minute-long enhancement in the solar wind dynamic pressure (1st red arrow, Fig \ref{tearing}a) triggered a two-hour-long thinning (1st red arrow, Fig \ref{tearing}c) and stretching (red arrow, Fig \ref{tearing}b) of the cross-tail current sheet. Thinning and stretching are important precursors for magnetic reconnection. During the preconditioning the current sheet thinned from its initial half-thickness $h$ of 1-to-2 Earth radii ($R_E=6378$ km) down to a few ion inertial lengths, based on the ion number density in the plasma sheet boundary layer ($d_{i0}\approx$860 km), and stretched to the point where the north-south magnetic field component $B_Z$ was less than 1\% of the magnetic field in the plasma sheet boundary layer ($B_{0}\approx$ 24 nT). The arrival of the second pressure pulse precipitated the rapid collapse of the current sheet (2nd red arrow, Fig \ref{tearing}c), the rapid growth of ion non-gyrotropy, and unloading of the magnetotail total pressure (see Paper 1). A strong and rapidly varying normal magnetic field component $B_Z$ and the formation and reversal of an ion reconnection exhaust jet was observed during the rapid collapse (blue line, Fig \ref{tearing}e); these signatures roughly mark the onset of magnetotail reconnection. The exact time and nature of the reconnection onset is to be determined in the next sections of this paper.

The criterion for the growth of the electron tearing instability is shown in Figure \ref{tearing}d, and is taken from \citeA{Liu.2014} to be 

\begin{equation}
2\frac{B_Zh}{B_0d_i}\sqrt{\frac{m_iT_i}{m_eT_e}}\leq1,
\end{equation}

\noindent where $m$ is the mass, $T$ is the temperature, and the subscripts ``e'' and ``i'' denote electron and ion quantities, respectively. The stretching (reduction in $B_Z/B_0$) and thinning (reduction in $h/d_i$) of the current sheet reduced the left-hand-side of equation 1 by two orders of magnitude during the preconditioning interval. Additionally, ion cooling resulting from the loss of high-energy plasma sheet ions (see paper 1) contributed to the reduction of the left-hand-side of equation 1; however, the reduction of $T_i/T_e$ ($\sim-30\%$ of initial value) between the first and second solar wind pressure pulses was very small compared to the reductions in $B_Z/B_0$ ($\sim-99\%$) and $h/d_i$ ($\sim-98\%$). The growth of the ion reconnection exhaust occurred during the rapid current sheet collapse, as the criterion for electron tearing was finally satisfied. We note that the current sheet thickness (and therefore the left-hand-side of equation 1) is likely overestimated during and after the rapid collapse stage, as $h$ was estimated using a single-layer Harris equilibrium, which does not take into account the nested thin current sheet configuration that is observed during reconnection onset \cite{Sitnov.2019b}. 

The remainder of this paper focuses on the several minutes before and during the reconnection onset period. 

\subsection{Rapid collapse of the current sheet}

The current sheet was very active during the rapid collapse and reconnection onset. Each mode of activity that is key to our story is detailed in a following subsection.

\begin{figure}
\noindent\includegraphics[width=42pc]{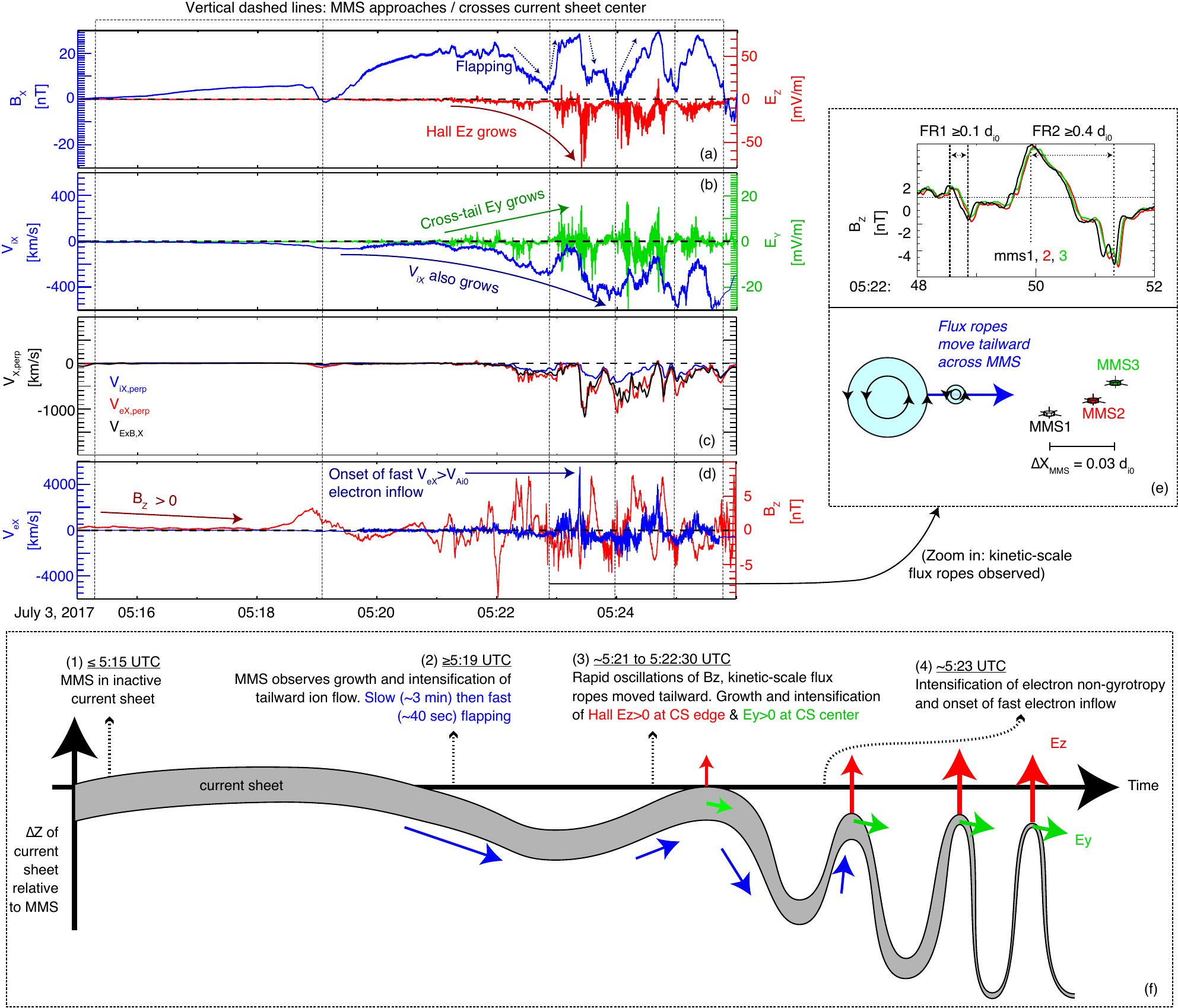}
\caption{Dynamics observed by MMS during reconnection onset. (a) The reconnecting component of the magnetic field (blue) and Hall electric field (red), (b) the outflow component of the ion velocity (blue) and cross-tail electric field (green), (c) the perpendicular components of the ion, electron, and $E\times B$ outflow velocities (blue, red, black, respectively), (d) the outflow component of the electron velocity (blue) and normal component of the magnetic field (red), (e) MMS data and a schematic diagram during encounters with flux ropes, (f) a schematic diagram illustrating the time history of events observed by MMS during the onset.}
\label{onset_timing}
\end{figure}

\subsubsection{Flapping wave growth}
Flapping is the periodic north-south motion of the current sheet, which can be seen in Figure \ref{onset_timing}a as oscillations in $B_X$. The magnitude of $B_X$ is a rough indicator of the distance between MMS and the current sheet center, as $B_X=0$ at the current sheet center and increases in magnitude with distance normal to the center. Flapping started as a longer-period wave around 5:15 UTC and subsequently increased in frequency, which allowed MMS to sample the vertical profile of the current sheet during the rapid current sheet collapse. Care must be taken to separate time dependencies in the MMS data associated with the evolution of the current sheet and time dependencies associated with motion of the current sheet, as some signatures of reconnection onset are unique to the plasma sheet boundary layer, the current sheet center, etc. Each flapping oscillation can therefore be considered a snapshot of the vertical profile of the current sheet and the current sheet evolution is evaluated from one snapshot to the next. Figure \ref{onset_timing}f illustrates the time history of events observed by MMS during each snapshot.  

\subsubsection{Embedded thin current layer formation}
Current sheet thinning is the concentration of the cross-tail current into one or multiple embedded layers. Embedded kinetic-scale current sheet formation occurs when the magnetic gradient scale length approaches or becomes smaller than the ion gyroradius. Speiser-type meandering motion of ions then occurs in a thin layer around the current sheet center and electrons $E\times B$ drift in the opposite direction. The thin current layer is embedded within the diamagnetic current of the broader plasma sheet. Paper 1 demonstrated the growth of the ion non-gyrotropy, which is one signature of ion meandering, accompanied the growth of the cross-tail current density (see Paper 1, Fig. 2m and 2o) during the rapid collapse stage. Since the motions of the meandering ions and drifting electrons are roughly in $Y$, perpendicular to $B_X$, one expects the growth of a Hall $E_Z$ ($\propto –V_YB_X$) that is between the current sheet center (where $B_X=0$) and the boundary of the thin current sheet (where $V_Y$ is small). The rapid growth of the Hall electric field is seen in Figure \ref{onset_timing}a. When MMS sampled the vertical profile of the current sheet around 5:20 UTC, no $E_Z$ was observed. The next three flapping intervals showed $E_Z$ grew exponentially from –10 mV/m to –80 mV/m in $\sim$1 minute. A strong $E_Z$ was then observed during every subsequent pass through the current sheet, which indicates the persistence of the embedded current layer. The differential drifts of ions and electrons, which is responsible for the $E_Z$ growth, also drives the growth of drift-kink-like flapping waves \cite{Daughton.1998}, so it makes sense that flapping accompanies thin current sheet formation. A quantitative analysis of the flapping instability is, however, beyond the scope of this study.

\subsubsection{Onset of electron tearing}
The electron tearing instability changes the magnetic topology in the current sheet. A hallmark of electron tearing is the interconnection of stretched magnetic field lines and the formation of magnetic flux ropes, in which the magnetic field has a wrapped, helical configuration. In a simplified, 2-d view, flux ropes resemble O-type magnetic field points and are separated by X-type points by necessity, as illustrated in Figure \ref{onset_timing}e; however, the topology in 3-d may be more complex. The passage of a flux rope across MMS is observed as a bipolar variation in $B_Z$. The clearest flux rope signatures begin appearing at MMS at $\sim$5:22:48 UTC, which roughly coincides with the time at which the electron tearing criterion was first satisfied ($\sim$5:20 UT, Fig. \ref{tearing}d). Two flux ropes are shown in (Fig. \ref{onset_timing}e). Differences in the timing of bipolar signatures at two of the spacecraft can be used to deduce linear speed. The product of the speed and the signal duration (measured from one $B_Z$ peak to the next) yields a lower bound of the flux rope size. This is a lower bound as the path MMS through the flux rope need not pass through the flux rope center. A delay of 40 milliseconds is observed between the MMS1 and MMS2 signals, while MMS 2 and 3 observed the flux ropes nearly simultaneously. Using the 40 ms delay and the 14 km separation between MMS1 and 2 in $X_{GSM}$ (assumed to be the propagation direction), the first flux rope is $\geq$90 km ($\geq 0.1d_i$) and the second flux rope is ($\geq 0.4d_i$). 

The flux ropes appear at MMS during the exponential growth of the Hall $E_Z$. During this time, MMS also observed the growth of sub-Alfv\'enic, tailward ($V_{iX}<0$) ion flow (Fig. \ref{onset_timing}b), where the ion Alfv\'en speed was estimated to be $\sim$400 km/s based on the background density and $B_X$ at the plasma sheet boundary. The ion flow remained tailward for the interval shown in Fig. \ref{onset_timing}, indicating the formation of a primary X-line that remained Earthward of MMS. The perpendicular component of the $V_{iX}$ was mostly smaller than the perpendicular electron ($V_{eX}$) and $E\times B$ speeds, where the latter two quantities were in good agreement with one another. The presence of these demagnetized ions, coupled with the sub-ion-scale thickness of the current sheet (Fig. \ref{tearing}c) and substantial ion non-gyrotropy (see Paper 1, Fig. 2o), may indicate that either (1) the ion population had not yet become fully entrained within the reconnection exhaust at this stage of the evolution or (2) MMS was located within a newly formed ion diffusion region.

The electron velocity $V_{eX}$ showed multiple reversals after 5:23 UT (Fig. \ref{onset_timing}d). Near the current sheet center, one expects the electron flow to be dominated by perpendicular convection driven by the reconnection exhaust. Away from the current sheet center, near the boundary of the exhaust, one expects an oppositely-directed parallel electron inflow. The parallel inflow and perpendicular outflow currents form the Hall current loop \cite{Sonnerup.1979}. Given that the perpendicular component of $V_{eX}$ remained negative throughout this interval, MMS likely sampled these different regions of the Hall current loop multiple times as the current sheet flapped up and down. After 5:23 UT, MMS observed an intensification of the parallel electron inflow, which exceeded the ion Alfv\'en speed. Given that the parallel electron inflow serves to balance the charge in the diffusion region, the intensification of $V_{eX}$ may be a remote signature of the intensification of the reconnection rate Earthward of MMS.

\subsection{Formation of a primary X-line}

\begin{figure}
\noindent\includegraphics[width=42pc]{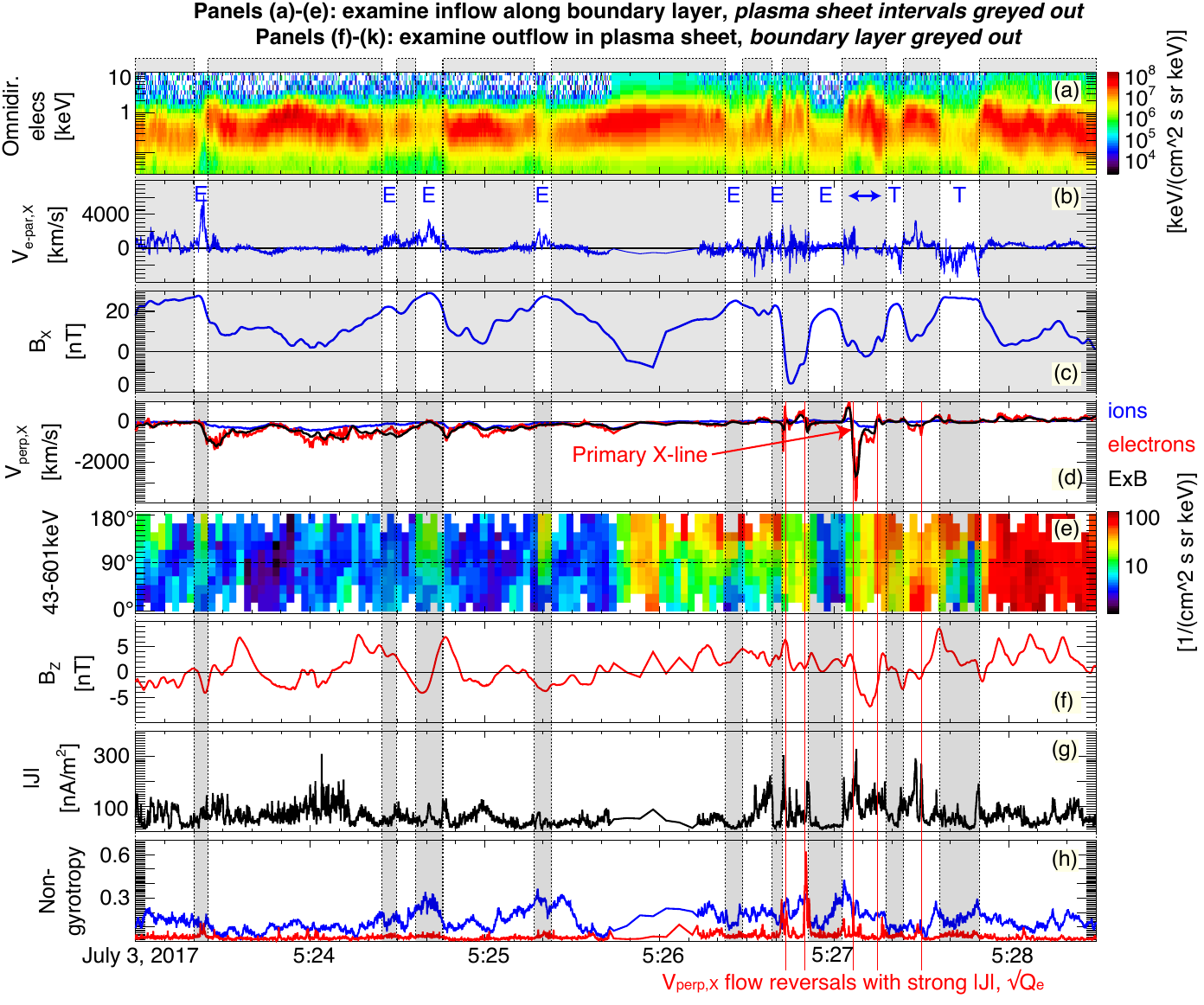}
\caption{(a) Omnidirectional energy fluxes, (b) field-aligned electron bulk velocity, (c) $B_X$, (d) the perpendicular outflow component of $V_i$ (blue), $V_e$ (red), and $V_{ExB}$ (black), (e) pitch angle spectrum of energetic (43-601 keV) electrons, (f) $B_Z$, (g) the magnitude of the current density, and (h) the ion (blue) and electron (red) non-gyrotropy, parameterized as $\sqrt{Q}$ \cite{Swisdak.2016}, where 0 indicates perfect gyrotropy and 1 is the maximum degree of non-gyroptopy.}
\label{primaryx}
\end{figure}

As shown in Fig. \ref{tearing}e, the ion outflow reversed direction once near 5:26 UT. Figure \ref{primaryx} zooms in on the interval around the flow reversal. If magnetotail reconnection onset involves the formation of multiple X-lines, then one expects to observe multiple flow convective reversals in the central plasma sheet and in the co-moving frame of each X-line. Similarly, one expects to observe multiple reversals of the parallel electron inflow in the outer plasma sheet. By definition, if reconnection is predominantly facilitated by a primary X-line, then all secondary reconnection X-lines are contained within the exhausts of the primary. Given that a primary X-line is distinguished from its secondary counterparts by its ability to access higher latitude lobe field lines, reversals of the electron inflows associated with primary X-lines should be observed at higher latitudes that those associated with secondary X-lines; these highest-latitude reconnecting field lines do not map directly to secondary diffusion regions, by definition. Therefore, a test of the primary/secondary X-line model requires the following observations (1) in the current sheet center MMS should observe multiple convective flow reversals but (2) in the outer boundary of the reconnection exhaust (the plasma sheet boundary layer) MMS should only observe one reversal of the parallel electron inflow (see for instance Fig. 1d-f in Paper 1).  Fig. \ref{primaryx}b-c show quantities used to diagnose the electron inflow and intervals when MMS is in the reconnection exhaust are greyed out. Fig. \ref{primaryx}d-h show quantities used to diagnose the outflow, and intervals when MMS is in the exhaust boundary are greyed out. The reconnection exhaust boundary is delineated from the current sheet center as the former has smaller electron fluxes (Fig. \ref{primaryx}a) and larger $B_X$ (Fig. \ref{primaryx}c) than the latter.

First we focus on the electron inflow. Fig. \ref{primaryx}b shows the electron parallel bulk flow velocity. Given that MMS is in the northern hemisphere ($B_X>0$), positive quantities indicate Earthward moving electrons and vice versa. For each encounter with the exhaust boundary, `E' and `T' are used to denote whether the electron motion is predominantly Earthward or tailward. Consistent with the primary X-line model, only one reversal of the electron inflow is observed between 5:27:00 and 5:27:30 UT.

Next we focus on the electron outflow. Since the electron exhausts are super-ion-Alfv\'enic in the diffusion region \cite{Phan.2007} and the typical speed of X-line motion is on the order of a tenth of the ion Alfv\'en speed \cite{Oka.2008,Alexandrova.2015}, reversals of $V_{eX\bot}$ can be used to identify reconnecting X-line crossings without needing to transform into the co-moving X-line frame. Five intervals, each marked with vertical red lines, are identified as possible X-line crossings by virtue of having (1) reversals in the convective electron motion $V_{eX\bot}$ (Fig. \ref{primaryx}d), (2) strongly enhanced current densities (Fig. \ref{primaryx}g), and (3) enhanced electron non-gyrotropy (Fig. \ref{primaryx}h). Of these five events, three (5:26:42, 5:26:50 and 5:27:07 UT) have been previously identified as electron diffusion regions (EDRs) \cite{Chen.2019}. Of the three EDRs, the event at 5:27:07 UT had the most pronounced electron outflows (Fig. \ref{primaryx}d) and by far the strongest reconnection electric field, being roughly five times larger than that of the other two events \cite{Chen.2019}. The EDR at 5:27:07 UT is also observed between the two exhaust boundary crossings where the electron inflow reversed, indicating that it may be the primary reconnection X-line.

One complication with the EDR at 5:27:07 is the sign of the electron perpendicular flow reversal; namely, MMS observed an Earthward-to-tailward $\vec{v}_{eX\bot}$, which is inconsistent with (1) the sign of the overall tailward-to-Earthward ion flow reversal and (2) the sign of the Earthward-to-tailward electron inflow reversal. Both (1) and (2) indicate an overall tailward motion of the primary X-line, which should be manifest in the data as a tailward-to-Earthward $\vec{v}_{eX\bot}$ reversal. Several possible explanations exist: first, the primary X-line may have a ``jittery'' back-and-forth motion, meaning that the EDR at 5:27:07 may have been encountered during a brief period of Earthward motion during its overall tailward retreat. Analysis of the electron distribution functions around flow reversal at 5:27:07 UT show strong, non-gyrotropic crescent-like features \cite{Chen.2020}, which is a key indicator that this event is likely an EDR \cite{Burch.2016b,Torbert.2018}. Alternatively, the crossing at 5:27:07 may have been a tailward-moving O-line bounded by two actively reconnecting X-lines. Field and flow reversal signatures look nearly identical for Earthward-moving X-lines and tailward-moving O-lines, which are best delineated by determining the direction of motion of the structure \cite{Eastwood.2010}. Unfortunately, timing analysis was inconclusive. MMS-1 observed the $B_Z$ reversal 0.02 seconds before MMS-3, which would indicate a tailward-moving O-line if the structure motion was purely along $X_{GSM}$ (following the MMS configuration in Fig. \ref{onset_timing}e). However, MMS-3 observed the $\vec{v}_{eX\bot}$ reversal 0.3 seconds before MMS-1, which would indicate an Earthward-moving X-line for the same assumptions. Full 3-d timing analysis, structure velocity determination with, e.g., the spatiotemporal differences method \cite{Shi.2006}, and reconstructions of the magnetic field structure \cite{Torbert.2020,Denton.2020} are not possible due to the lack of MMS-4 data. Still, and regardless of whether the 5:27:07 UT event was an X-line or an O-line bounding active X-lines, the conclusion remains that the primary X-line was very near 5:27:07 UT.

\subsection{Energetic electrons and the magnetic topology}

A strong enhancement in the flux of energetic electrons was observed over all pitch angles after the primary EDR encounter (Fig. \ref{primaryx}e). Prior to the primary EDR encounter, substantial energetic electron fluxes were only detected in the anti-field-aligned (tailward) direction. The marked change in the distribution of energetic electrons is consistent with the transition of MMS from open field lines in the tailward exhaust, where electrons can stream freely into the deep tail environment, to closed field lines in the Earthward exhaust, where trapped electrons bounce back and forth along field lines as they mirror between the magnetic poles. If the change in the electron spectrum is indeed indicative of a change in the magnetic topology (open to closed field lines) then these data lend further credence to the conclusion that reconnection was occurring at a primary X-line observed at or near 5:27:07 UT.

\section{Summary and conclusions}

The local dynamics of a magnetotail reconnection onset event were investigated with MMS data. The event was characterized by rapid current sheet thinning, the onset of current sheet flapping and subsequent onset of electron tearing, and the formation of multiple kinetic-scale flux ropes and reconnecting X-lines. The current sheet thinning was accompanied by the growth of a Hall electric field. The development of a primary X-line was remotely detected by MMS as the growth of an initially sub-Alfv\'enic, demagnetized ion jet in the current sheet center followed by the sudden enhancement of super-Alfv\'enic electron inflow at the exhaust boundary. 

The single reversal of the spacecraft-frame ion outflow velocity was associated with the formation of a primary X-line. The primary X-line had access to, and reconnected, higher latitude field lines than secondary X-lines entrained within its exhaust. This point was evidenced by the identification of a single reversal in the electron inflow at the exhaust boundary coincident with an EDR in the current sheet center. No such inflow reversal was found to be associated with two additional EDRs, both of which were previously found to have slower reconnection rates than the primary EDR. Strong enhancements in the fluxes of energetic electrons were observed in the vicinity of the primary X-line. Tailward of the primary X-line, energetic electrons were observed as an anti-parallel streaming population, whereas a broad isotropic distribution was observed Earthward of the primary X-line. This change in the distribution is consistent with the magnetic topology changing due to reconnection at the primary X-line.

\acknowledgments
KJG, CJF, and RBT were supported by NASA’s MMS FIELDS contract NNG04EB99C. 

\section{Open Research}
MMS data, including FGM \cite{MMS_FGM_BRST}, FPI \cite{MMS_DES_MOMS_BRST,MMS_DIS_MOMS_BRST}, EPD \cite{MMS_EPD_BRST}, and FEEPS \cite{MMS_FEEPS_BRST}, are publicly available were obtained from https://lasp.colorado.edu/mms/sdc/public. This study utilized the Space Physics Environment Data Analysis System (SPEDAS) software package \cite{spedas}.


%
%


%
%
%
%
%

\end{document}